\documentclass[nofootinbib]{revtex4}

\usepackage[english]{babel}
\usepackage{graphicx}
\usepackage{psfrag}
\usepackage{amsmath}
\usepackage{amssymb}
\usepackage{verbatim}

\newcommand{\be}{\begin{equation}}
\newcommand{\ee}{\end{equation}}
\newcommand{\bea}{\begin{eqnarray}}
\newcommand{\eea}{\end{eqnarray}}
\newcommand{\bean}{\begin{eqnarray*}}
\newcommand{\eean}{\end{eqnarray*}}

\renewcommand{\b}{\langle}
\newcommand{\ket}{\rangle}

\newcommand{\irm}{{\rm i}}
\newcommand{\e}{{\rm e}}

\renewcommand{\d}{{\rm d}}
\newcommand{\cl}[1]{{\mathcal #1}}

\newcommand{\pa}{\partial}

\newcommand{\ts}{\textstyle}
\newcommand{\sst}{\scriptstyle}

\newcommand{\pdiff}[2]{\frac{\partial #1}{\partial #2}}

\newcommand{\bZ}{\mathbb{Z}}
\newcommand{\bN}{\mathbb{N}}

\newcommand{\bR}{\mathbb{R}}

\newcommand{\clS}{\cl{S}}

\newcommand{\eq}[1]{(\ref{#1})}
\renewcommand{\sec}[1]{sec.\ \ref{#1}}

\newcommand{\fig}[1]{Fig.\ \ref{#1}}

\newcommand{\tr}{{\rm tr}}

\newcommand{\pic}[4]
{
 \begin{figure}
 \begin{center}
 \includegraphics[height=#3]{#4}
 \end{center}
 \caption{\label{#1} #2}
 \end{figure}
}

\newcommand{\qed}{\nobreak \ifvmode \relax \else
      \ifdim\lastskip<1.5em \hskip-\lastskip
      \hskip1.5em plus0em minus0.5em \fi \nobreak
      \vrule height0.75em width0.5em depth0.25em\fi}

\newcommand{\muh}{\hat{\mu}}

\newcommand{\bb}{\overline{b}}

\newcommand{\nablab}{\overline{\nabla}}
\newcommand{\xb}{\overline{x}}

\newcommand{\tDel}{\tilde{\Delta}}

\begin{document}
\thispagestyle{empty}
\hfill
\parbox[t]{3.6cm}{
hep-th/yymmnnn \\
IGPG-06/10-6} 
\vspace{2cm}

\title{Analytic derivation of dual gluons and monopoles \\ from SU(2) lattice Yang-Mills theory \\ III.\ Plaquette representation}
\author{Florian Conrady}
\affiliation{Institute for Gravitational Physics and Geometry, Physics Department, Penn State University, University Park, PA 16802, U.S.A}
\email{conrady@gravity.psu.edu}
\preprint{IGPG-06/10-6}

\begin{abstract}
In this series of three papers, we generalize the derivation of dual photons and monopoles by Polyakov, and Banks, Myerson and Kogut,
to obtain approximative models of SU(2) lattice gauge theory. Our approach is based on stationary phase approximations.

In this third article, we start from the plaquette representation of 3-dimensional SU(2) lattice gauge theory. 
By extending a work of Borisenko, Voloshin and Faber, we transform the expectation value of a Wilson loop into a path integral
over a dual gluon field and monopole variables. The action contains the tree-level Coulomb interaction
and a nonlinear coupling between dual gluons, monopoles and current.

We show that this model exhibits confinement if a condition on the monopole self-energy is satisfied.
\end{abstract}

\maketitle

\section{Introduction}
\label{introduction}

In the analysis of lattice gauge theories, it proves very useful to transform between different representations of the same theory. In the 70's, 
Banks, Myerson, and Kogut have shown that U(1) lattice gauge theory can be transformed exactly into a representation by a dual photon field and monopoles \cite{BanksMyersonKogut}. This photon-monopole representation was derived earlier by a different method by Polyakov \cite{PolyakovI,PolyakovII}.

One can demonstrate with it that electrostatic charges are confined in 3 dimensions \cite{PolyakovII,GopfertMack}, and that there is a phase transition in 4 dimension \cite{BanksMyersonKogut,Guth,FrohlichSpencer}: in $d=3$ the monopoles condense along a string between the charges and create a linear confining potential. In dimension 4 there is a critical size of the coupling at which such a condensation sets in.

The example of U(1) fostered hopes that one could generalize this scheme to non-abelian gauge groups: it led to the conjecture that confinement in U(1) and SU(N) result from the same mechanism, and that it has an analogy in dual type II superconductors \cite{tHooftsuperconductor,Mandelstam}. 

The problem, however, proves to be much harder than in the abelian case. Presently, we only know of three non-abelian analogues of U(1) lattice representations: apart from the defining path integral, we have 1.\ a first-order representation, which can be viewed as a lattice version of BF Yang-Mills theory, 2.\ the spin foam representation \cite{Anishettyetal,HallidaySuranyi,OecklPfeifferdualofpurenonAbelian,Conradygeometricspinfoams}, and 3.\ the plaquette representation \cite{Batrouni,BatrouniHalpern,BorisenkoVoloshinFaber}. The first two representations are defined for any dimension $d\ge 2$, while the latter has been only constructed in 3 dimensions so far. 

It is difficult to go beyond these three representations: both the plaquette and the spin foam representation are governed by non-abelian generalizations of constraints that we find in the abelian case. For U(1) these constraints can be solved exactly, so that one can move on to obtain the dual $\bZ$ gauge theory and the photon-monopole representation. In the non-abelian case, we do not know at present, how one could solve these constraints exactly, or if there is at all a meaningful sense in which representations beyond the known ones exist. 

This article is the third in a series of three papers, where we derive non-abelian analogues of the 
photon-monopole representation for gauge group SU(2). We are not able to do this by exact transformations, and have to rely instead on stationary phase approximations. This leads us to three versions of a gluon-monopole representation of SU(2) lattice gauge theory \cite{ConradyglumonI,ConradyglumonII}. In the present paper, our starting point is the plaquette representation in 3 dimensions, as it was given by Borisenko, Voloshin and Faber\footnote{It is related to earlier plaquette (or field strength) representations by Halpern and Batrouni \cite{Halpern,Batrouni,BatrouniHalpern}.}. These authors developed a weak-coupling perturbation theory for the plaquette representation, and derived the associated generating functional. It is a path integral over continuous and discrete variables that resemble the dual photons and monopoles of U(1). We interpret these variables as a dual gluon and monopole field of SU(2). 

We extend the work of Borisenko et al.\ by including a Wilson loop as a non-trivial background of the perturbation theory: we consider the expectation value of a Wilson loop, and treat the latter in such a way that it contributes at zeroth order. This is done with the help of the Kirillov trace formula. We then retain only the zeroth order and arrive at a gluon-monopole model that has a coupling to a source current $J$. The structure of the resulting action is analogous to the photon-monopole action: it reproduces roughly the tree-level Coulomb interaction, and the coupling between monopoles, dual gluons and current is similar to the abelian case. There is an important difference, however: it consists in the fact that the dual gluon field is $\mathrm{su(2)}\simeq \bR^3$-valued (and not $\bR$-valued) and that the monopoles couple to the \textit{length} of field vectors. As a result, the gluon-monopole coupling is nonlinear.

The similarity to U(1) naturally suggests that one could try to generalize Polyakov's method of deriving confinement. We propose a way to do so, but it requires an additional heuristic assumption: since we are not able to integrate over the dual gluon field as it was done for U(1), we have to \textit{assume} a renormalization group step that generates a monopole self-energy at a scale $M$ below the cutoff scale. With that assumption, the rest follows as in Polyakov's paper and one arrives at a non-vanishing string tension for the Wilson loop in the representation $j=1/2$.

The paper is organized as follows: the definition of the plaquette representation is given in \sec{plaquetterepresentationof3dimensionalSU2latticeYangMillstheory}. 
In \sec{representationasgluonsandmonopoles}, we derive the approximative representation by dual gluons and monopoles. The tentative derivation of the area law follows in \sec{derivationofthearealaw}. The results are summarized and discussed in the final section.

\subsubsection*{\textit{Notation and conventions}}

$\kappa$ denotes a 3-dimensional hypercubic lattice of side length $L$ with periodic boundary conditions. The lattice constant is $a$. 
Depending on the context, we use abstract or index notation to denote oriented cells of $\kappa$: in the abstract notation, vertices, edges, faces and cubes are written as $v$, $e$, $f$ and $c$ respectively. In the index notation, we write $x$, $(x\mu)$, $(x\mu\nu)$, $(x\mu\nu\rho)$ etc. Correspondingly, we have two notations for chains. Since the lattice is finite, we can identify chains and cochains. As usual, $\pa$, $\d$ and $*$ designate the boundary, coboundary and Hodge dual operator respectively. Forward and backward derivative are defined by
\be
\nabla_\mu f_x = \frac{1}{a}\left(f_{x+a\muh} - f_x\right)\,,\qquad \nablab_\mu f_x = \frac{1}{a}\left(f_x - f_{x-a\muh}\right)
\ee
where $\muh$ is the unit vector in the $\mu$-direction. The lattice Laplacian reads
\be
\Delta = \nablab_\mu \nabla_\mu\,.
\ee
For a given unit vector $u = \muh$ and a 1-chain $J_{x\mu}$, we define
\be
(u\cdot\nabla)^{-1} J_{x\mu} := \sum_{x'_\mu \le x_\mu} J_{(x_1,\ldots,x'_\mu,\ldots,x_3)\,\mu}\,.
\ee
Color indices are denoted by $a, b, c, \ldots$ We employ units in which $\hbar = c = 1$ and $a = 1$. For some quantities, the $a$-dependence is indicated explicitly.

\section{Plaquette representation of 3-dimensional SU(2) lattice Yang-Mills theory}
\label{plaquetterepresentationof3dimensionalSU2latticeYangMillstheory}

\pic{eightcubes}{Eight cubes around an even point: the thick lines indicate a possible choice of connectors.}{5cm}{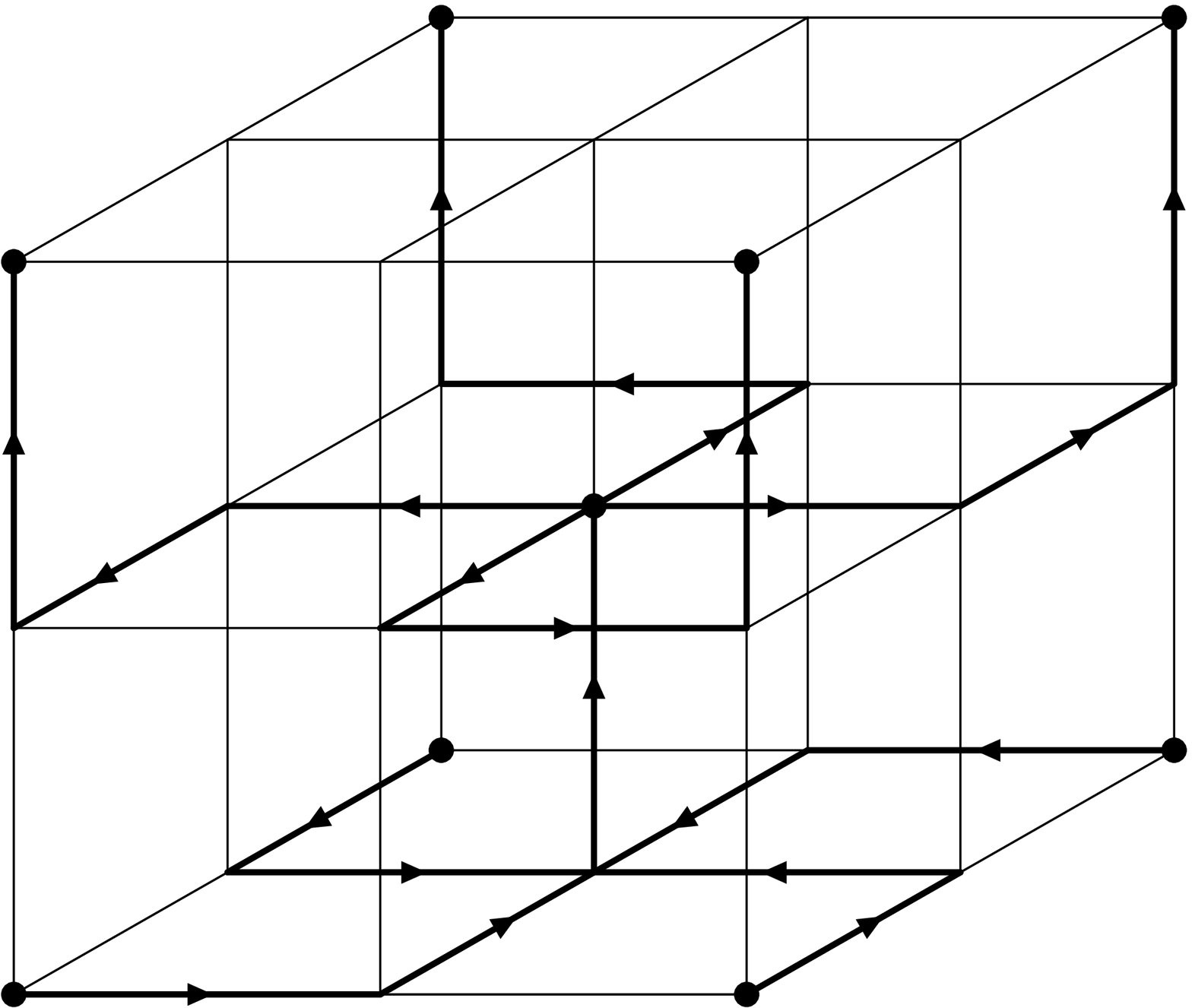}

The basic idea behind the plaquette (or field-strength) formulation is a change of variables
from holonomies along edges to holonomies around plaquettes. The path integral over the new plaquette 
variables is constrained, since plaquette holonomies have to satisfy Bianchi constraints: 
the well-known Bianchi identity of abelian lattice gauge theory, or generalizations thereof for 
non-abelian gauge theories.

There exist different schemes for constructing such plaquette representations (see \cite{Batrouni,BatrouniHalpern,BorisenkoVoloshinFaber}).
In this paper, we will use the formulation of Borisenko, Voloshin and Faber for
SU(N) lattice gauge theory in 3 dimensions \cite{BorisenkoVoloshinFaber}.
Although the underlying idea is simple, a precise description of this representation is quite involved,
as it requires numerous conventions on orientations and orderings. 
For that reason, we will review the essential definitions in this section.

To simplify the presentation, we will ignore all boundary effects. In the next section, we will restrict ourselves to
the first term in a weak-coupling expansion, and to that order any boundary-related modifications drop out.
The proper treatment of boundary conditions is given in \cite{BorisenkoVoloshinFaber}.

Edge (or link) variables are denoted by $U_e$, and $W_f$ designates the holonomy around a face $f$:
\be
\label{constraintonWf}
W_f = {\ts\prod\limits_{e\subset\pa f}} U_e
\ee
It is assumed that we have chosen some starting point $v\subset f$, so that $W_f$ is the product of edge holonomies $U_e$, starting at $v$, and following the orientation of the face $f$. Whenever we have such (or similar) products of group elements, we use left multiplication, i.e.\ 
\be
\prod_{i=0}^n U_i = U_n U_{n-1}\ldots U_2 U_1\,.
\ee
The convention for starting points will be fixed further below.

In the standard formulation of lattice gauge theory, the expectation value of a Wilson loop $C$ is defined by the path integral
\be
\label{standardformulation}
\b\tr_j W_C\ket = \int\left({\ts\prod\limits_{e\subset\kappa}}\;\d U_e\right)
\exp\left[-\sum_{f\subset\kappa}\;\frac{\beta}{4}\,\tr\left(W_f + W^{-1}_f\right)\right]\chi_j\left(W_C\right)\,.
\ee
Here, the exponent is given by the Wilson action and 
\be
\beta = \frac{4}{a g^2}\,.
\ee
$W_C$ stands for the holonomy around the Wilson loop $C$, where, again, a starting point along $C$ is assumed.
$\chi_j = \tr_j$ is the character in the representation $j$. It is understood that for each pair of edge orientations $e$, $e^{-1}$, we integrate only over one edge variable $U_e$, with the other one being fixed by
\be
U_{e^{-1}} = U^{-1}_e\,.
\ee

The plaquette representation arises from a change of variables: from the edge (or link) variables $U_e$ to the $W_f$'s, which we call face (or plaquette)
 variables. This change of variables is achieved in four steps:
\begin{enumerate}
\item Introduce integrations over plaquette variables $W_f$ + delta constraints \eq{constraintonWf} on them.
\item Reexpress some of these constraints as non-abelian Bianchi identities.
\item Impose the maximal axial gauge.
\item Integrate out the edge variables $U_e$.
\end{enumerate}
The result is a path integral over plaquette variables where of all the constraints on $W_f$ only one type is left, namely, the non-abelian Bianchi identities.

\psfrag{1}{$\sst 1$}
\psfrag{2}{$\sst 2$}
\psfrag{3}{$\sst 3$}
\psfrag{1'}{$\sst 1'$}
\psfrag{2'}{$\sst 2'$}
\psfrag{3'}{$\sst 3'$}
\psfrag{v}{$\sst v$}
\psfrag{v'}{$\sst v'$}
\psfrag{e1}{$\sst e_1$}
\psfrag{e2}{$\sst e_2$}
\psfrag{e3}{$\sst e_2$}
\pic{cubeconnector}{Cube of $\kappa$ with chosen connector.}{4cm}{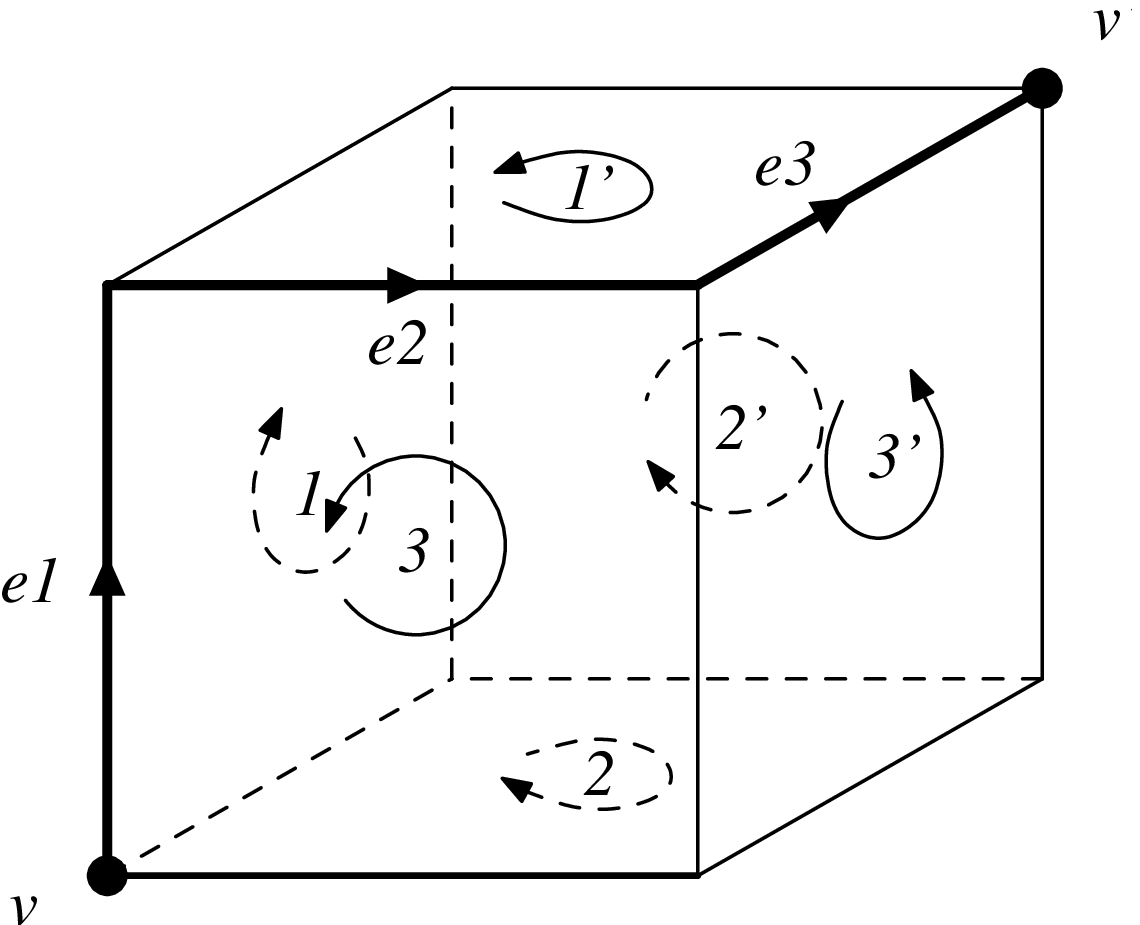}

To write down the non-abelian Bianchi identities we need to introduce a number of conventions: 
consider the set of vertices $v$ whose coordinates $(x_1,x_2,x_3)$ are all even and call them the even vertices.
Likewise, we call the vertices with odd coordinates odd. Clearly, each cube of $\kappa$ contains an even and odd point as corners, and each face of $\kappa$ contains either one even or one odd vertex. We will use this to specify our convention for starting points of face holonomies $W_f$: when associating a holonomy to a face we will always take the even or odd point in it as the starting point. 
 
In every cube of $\kappa$ we choose an oriented path consisting of three edges that connect the even and odd point. We call this path a connector. We can choose the connectors in such a way that there are altogether four types of connectors on the lattice (see \fig{eightcubes}). To each connector we associate a holonomy
\be
K_{v'v} = U_{e_3} U_{e_2} U_{e_1}\,.
\ee
Let us now specify the non-abelian Bianchi identities. Consider one cube $c$ as in \fig{cubeconnector}. Take the boundary $\pa c$ and choose an orientation for it. This will induce an orientation for each face $f\subset\pa c$. Call the starting point of the connector $v$ and the end point $v'$. 
We have three faces of the cube that meet at $v$: order them $f_1, f_2, f_3$, starting with the face whose orientation agrees with that of the connector, then the one which does not intersect with the connector, and then the remaining one (whose orientation is opposite to that of the connector). Likewise, we have three faces meeting at $v'$, and we order them $f'_1, f'_2, f'_3$, according to the same rule as before (see \fig{cubeconnector}).  

With these conventions the non-abelian Bianchi identity looks as follows:
\be
\label{Bianchiconstraint}
V_c = K^{-1}_{p'p}\,W_{f'_3}W_{f'_2}W_{f'_1}\,K_{p'p}\,W_{f_3}W_{f_2}W_{f_1} = I
\ee
$W_{f_i}$ denotes the holonomy around the face $f_i$ with the starting point at $p$, and likewise the starting point for $W_{f'_i}$ is taken to be $p'$.
The reader may convince himself by a drawing that this identity is indeed satisfied.

In the abelian case, the connectors would cancel and the expression would reduce to the usual abelian Bianchi identity. Here, this does not happen.
At first sight one might think that this makes a formulation in terms of plaquette variables impossible, since the appearance of connectors
prevents us from eliminating the edge variables completely. It turns out, however, that by the steps 3 and 4 above, the edge variables appearing in \eq{Bianchiconstraint} will be expressed in terms of plaquette variables, and thus one ends up with a path integral that contains only plaquette variables plus constraints on them.

\setlength{\jot}{0.3cm}
The expressions for the edge variables depend on how the edge is situated relative to the origin and the even and odd points: 
for example, if $x$ is an even point and $x_i > 0$, the edge variable $U_{x1}$ is given by (see \fig{edgevariable})
\be
\label{exampleedgevariable}
U_{(x_1,x_2,x_3),1} = \prod_{x'_2=x_2-1}^0 W_{(x_1,x'_2,0),21} \prod_{x'_3=x_3-1}^0 W_{(x_1,x_2,x'_3),31}\,.
\ee
Similar equations hold for other edge variables. To write them down explicitly, we would have to distinguish many different cases, although the principle is the same as in equation \eq{exampleedgevariable}. In this paper, these details do not matter, since we only use the zeroth order of a weak-coupling expansion, and the formulas for the connectors and edge variables drop out.

\psfrag{x1}{$\sst x_1$}
\psfrag{x2}{$\sst x_2$}
\psfrag{x3}{$\sst x_3$}
\psfrag{x}{$\sst x$}
\psfrag{Ux1}{$\sst U_{x1}$}
\pic{edgevariable}{Edge variables as products of plaquette variables: in the maximal axial gauge the edge variable $U_{x1}$ can be expressed as a product of plaquette variables. The dots represent even points, and the arrows indicate the order of the product. All edge variables on dashed lines equal $I$ due to the maximal axial gauge.}{5cm}{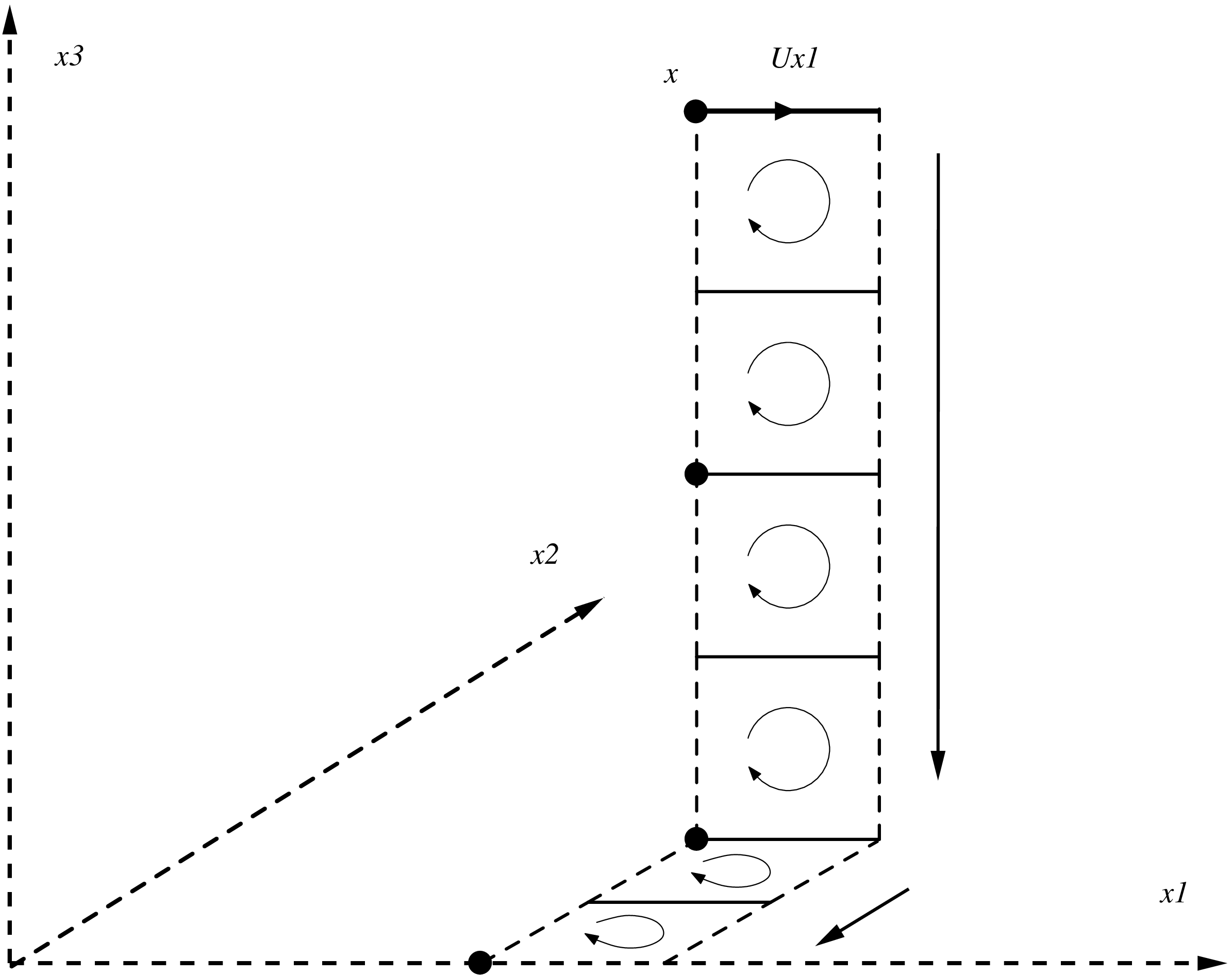}

In a similar fashion the Wilson loop can be expressed as a product of plaquette variables. We take $C$ to be a rectangular loop in the 1-2-plane with anti-clockwise orientation, where the lower-left and upper-right corner are given by even points $(0,0,0)$ and $(T,R,0)$, $T, R>0$. Then, the holonomy around $C$ with starting point $x = (0,0,0)$ equals
\be
\label{WilsonloopintermsofWs}
W_C = \prod_{x_1=T-1}^0\;\prod_{x_2=R-a}^{0} W_{(x_1,x_2,0),12}\,,
\ee
i.e.\ it is an ordered product of plaquette variables whose plaquettes fill the surface enclosed by the Wilson loop.

By going through the steps 1 to 4 and using the formulas \eq{Bianchiconstraint}, \eq{exampleedgevariable} and \eq{WilsonloopintermsofWs}, the original path integral \eq{standardformulation} is rewritten as a constrained path integral over plaquette variables $W_f$:
\be
\label{plaquetterepresentation}
\b\tr_j W_C\ket = 
\int\left({\ts\prod\limits_{f\subset\kappa}}\;\d W_f\right)
\left({\ts\prod\limits_{c\subset\kappa}}\;\delta\left(V_c\right)\right)
\exp\left[\sum_f\;\frac{\beta}{4}\,\tr\left(W_f + W^{-1}_f\right)\right]\chi_j\left(W_C\right)
\ee
If we change to an index notation, writing $(x\mu)$ instead of $e$ and $(x\mu\nu)$ in place of $f$, the same reads
\be
\b\tr_j W_C\ket = 
\int\left({\ts\prod\limits_{x\mu\nu}}\;\d W_{x\mu\nu}\right)
\left({\ts\prod\limits_x}\;\delta\left(V_x\right)\right)
\exp\left[\sum_x\sum_{\mu < \nu}\;\frac{\beta}{4}\,\tr\left(W_{x\mu\nu} + W^{-1}_{x\mu\nu}\right)\right]\chi_j\left(W_C\right)\,.
\ee

\section{Representation as dual gluons and monopoles}
\label{representationasgluonsandmonopoles}

After deriving their plaquette representation, Borisenko et al.\ continue with the construction of a weak-coupling (large $\beta$) 
perturbation theory. The associated generating functional can be considered as a partition function of dual gluons and monopoles of SU(2) lattice gauge theory,
and generalizes the photon-monopole partition function of U(1).

What we add to this scheme is the following: we use the Kirillov trace formula \cite{Kirillov}, as in paper I, to incorporate the Wilson loop in the exponent.
With this step we find a generalization of the photon-monopole action that does not only contain the dual gluons and monopoles, 
but also the current of the Wilson loop. This will be crucial in section \ref{derivationofthearealaw} where we propose a non-abelian generalization of Polyakov's derivation of the area law.

We will now describe the steps that lead to the description in terms of dual gluons and monopoles: first we will rewrite measure, action and Bianchi constraint in terms of Lie algebra elements. To do this for the Bianchi constraint, the delta function is expressed as a sum over characters, and the character, by the Kirillov trace formula, as a function of a Lie algebra element. We apply the same formula to the trace over the Wilson loop, so that it is parametrized by a Lie algebra element. 

In the second step, we apply the Poisson summation formula which trades the discrete sum over representations for new degrees of freedom that can be regarded as dual gluons and monopoles. The third and final step is a stationary phase approximation.

\subsection*{Kirillov trace and Poisson summation formula}
\label{KirillovtraceandPoissonsummationformula}

We start by rewriting everything in terms of Lie algebra elements: if we parametrize the plaquette variables as
\be
W_{x\mu\nu} = \e^{\irm\,\omega^a_{x\mu\nu}\sigma^a/2}\,,\qquad |\omega_{x\mu\nu}| < 2\pi\,,
\ee
the action becomes
\be
\frac{\beta}{4}\,\tr\left(W_{x\mu\nu}+W^{-1}_{x\mu\nu}\right)
 = \frac{\beta}{2}\,\cos\left(|\omega_{x\mu\nu}|/2\right)\,. 
\ee
The measure takes the form
\be
\int\d W_{x\mu\nu}\;\ldots 
\quad = \quad
\frac{1}{\pi^2}\int\limits_{B_{2\pi}(0)}\d^3\omega_{x\mu\nu}\;
\frac{\sin^2\left(|\omega_{x\mu\nu}|/2\right)}{\left(|\omega_{x\mu\nu}|/2\right)^2}\;\ldots\,.
\ee
The Bianchi constraint is written as
\be
\delta\left(V_x\right) = \sum_j\,(2j+1)\,\chi_j(V_x)\,.
\ee
Let $\upsilon^a_x \sigma^a/2$ be the Lie algebra element associated to the constraint $V_x$, i.e.
\be
V_x = \e^{\irm\,\upsilon^a_x \sigma^a/2}
\ee
Then the Kirillov trace formula gives
\be
\chi_j\left(V_x\right) = 
\frac{(2j+1)|\upsilon_x|/2}{4\pi\sin(|\upsilon_x|/2)}\,\int_{S^2} \d n\;\,\e^{\irm\,(2j+1)\,n\cdot\upsilon_x/2}\,.
\ee
The integral runs over unit vectors $n$ in $\bR^3$, i.e.\ over the 2-sphere. Next we use the Poisson summation formula to replace the sum over representations by an integral \textit{and} a sum: \setlength{\jot}{0.3cm}
\bea
\sum_j\,(2j+1)\chi_j(V_x) &=& \sum_{k\in\bN_0}\;\frac{k^2|\upsilon_x|/2}{4\pi\sin(|\upsilon_x|/2)}\,\int_{S^2} \d n\;\,\e^{\irm\,k\,n\cdot\upsilon_x/2} \nonumber \\
&=& \frac{1}{2}\sum_{k\in\bZ}\;\frac{k^2|\upsilon_x|/2}{4\pi\sin(|\upsilon_x|/2)}\,\int_{S^2} \d n\;\,\e^{\irm\,k\,n\cdot\upsilon_x/2} \nonumber \\
&=& 
\frac{1}{2}\int_{-\infty}^\infty \d r\,r^2\;\sum_{m\in\bZ}\;\frac{|\upsilon_x|/2}{4\pi\sin(|\upsilon_x|/2)}\,\int_{S^2} \d n\;\,\e^{\irm\,r\,n\cdot\upsilon_x/2\;+\; 2\pi\irm\,r\,m} \nonumber \\
&=& \frac{1}{8\pi}\int\limits_{\bR^3}\d^3\varphi\;\sum_{m\in\bZ}\;\frac{|\upsilon_x|/2}{\sin(|\upsilon_x|/2)}\,\e^{\irm\,\varphi\cdot\upsilon_x\,+\; 4\pi\irm\,|\varphi|m}
\eea
In the last equation we combined the integral over the radius and the unit vector into an integral over $\bR^3$.

To deal with the Wilson loop, we apply the Kirillov trace formula a second time: by writing \setlength{\jot}{0cm}
\be
W_C  = \e^{\irm\,\omega^a_C \sigma^a/2}
\ee
we obtain
\be
\chi_j\left(W_C\right) = 
\frac{(2j+1)|\omega_C|/2}{4\pi\sin(|\omega_C|/2)}\,\int_{S^2} \d n\;\,\e^{\irm\,(2j+1)\,n\cdot\omega_C/2}\,.
\ee
With all this the path integral becomes \setlength{\jot}{0.3cm}
\bea
\label{pathintegralparametrization}
\lefteqn{\b \tr_j W_C\ket =
\frac{1}{Z}\,
\int\limits_{B_{2\pi}(0)}\left({\ts\prod\limits_{x\mu\nu}}\;\d^3\omega_{x\mu\nu}\right)
\int\limits_{\bR^3}\left({\ts\prod\limits_x}\;\d^3\varphi_x\right)\sum_{m_x\in\bZ}\;\;\int\limits_{S^2} \d n} \nonumber \\
&& \times\,
\left({\ts\prod\limits_x}\;\frac{|\upsilon_x|/2}{\sin(|\upsilon_x|/2)}\right)\;
\frac{(2j+1)|\omega_C|/2}{4\pi\sin(|\omega_C|/2)}\nonumber \\
&& \times\,
\exp\left[\sum_x\left(
\sum_{\mu < \nu}\beta\cos\Big(|\omega_{x\mu\nu}|/2\Big) + \irm\,(2j+1)\,n\cdot\omega_C/2 + \irm\,\varphi_x\cdot\upsilon_x + 4\pi\irm\,|\varphi_x|\,m_x
\right)\right] \nonumber \\
&&
\eea
The integration range of $\omega_{x\mu\nu}$ is the ball of radius $2\pi$ in $\bR^3$. The scalar fields $\varphi$ and $m$ are $\bR^3$- and $\bZ$-valued respectively. Field-independent constants that appear both in the numerator and in $Z$ were dropped.
\setlength{\jot}{0cm}

\subsection*{Stationary phase approximation}
\label{stationaryphaseapproximation}

Let us expand all quantities in \eq{pathintegralparametrization} in powers of $\omega_{x\mu\nu}$: the Wilson action yields
\be
\label{expansionaction}
\beta\cos\Big(|\omega_{x\mu\nu}|/2\Big) = \beta\left(1 - \frac{1}{16}\,\omega^2_{x\mu\nu} + o(\omega^3_{x\mu\nu})\right)\,.
\ee
When we expand the non-abelian Bianchi identity \eq{Bianchiconstraint}, the contribution of the connectors cancel to lowest order, and
\be
\label{expansionBianchiconstraint}
\upsilon_x = \frac{1}{2}\,\epsilon_{\rho\mu\nu} \nabla_\rho\omega_{x\mu\nu} + o(\omega^2_{x\mu\nu})\,.
\ee
It follows from equation \eq{WilsonloopintermsofWs} that
\be
\label{expansionWilsonloop}
\omega_C = \sum_x \frac{1}{2}\,S_{x\mu\nu}\omega_{x\mu\nu} + o(\omega^2_{x\mu\nu})\,,
\ee
where $S$ is the minimal surface spanned by the Wilson loop $C$. From the measure factors we obtain
\be
\label{expansionmeasure1}
\frac{|\omega_{x\mu\nu}|/2}{\sin(|\omega_{x\mu\nu}|/2)}
= \exp\left(\frac{1}{6}\left(\frac{|\omega_{x\mu\nu}|}{2}\right)^2 + \frac{1}{180}\left(\frac{|\omega_{x\mu\nu}|}{2}\right)^4 + \ldots\right)\,.
\ee
\be
\label{expansionmeasure2}
\frac{|\upsilon_x|/2}{\sin(|\upsilon_x|/2)} = \exp\left(\frac{1}{6}\left(\frac{|\frac{1}{2}\epsilon_{\rho\mu\nu} \nabla_\rho\omega_{x\mu\nu} + \ldots|}{2}\right)^2 + \ldots\right)\,.
\ee
\be
\label{expansionmeasure3}
\frac{|\omega_C|/2}{\sin(|\omega_C|/2)} = \exp\left(\frac{1}{6}\left(\frac{|\sum_x \frac{1}{2}\,S_{x\mu\nu}\omega_{x\mu\nu} + \ldots|}{2}\right)^2 + \ldots\right)\,.
\ee
Each of the terms \eq{expansionaction}--\eq{expansionmeasure3} contributes with linear, quadratic and higher orders of $\omega_{x\mu\nu}$.

We now apply a stationary phase approximation at $\omega_{x\mu\nu} = 0$ and only retain terms up to quadratic order in $\omega_{x\mu\nu}$.
We consider the continuum limit where $\beta \gg 1$: then, the quadratic term in \eq{expansionaction} is much larger than the quadratic terms coming from 
\eq{expansionBianchiconstraint}--\eq{expansionmeasure3}, so we keep only the former. This yields the following path integral\footnote{Throughout it is assumed that we apply the same steps within the partition function $Z$ by which we divide.}:
\bea
\lefteqn{\b \tr_j W_C\ket =
\frac{1}{Z}\,
\int\limits_{B_{2\pi}(0)}\left({\ts\prod\limits_{x\mu\nu}}\;\d^3\omega_{x\mu\nu}\right)
\int\limits_{\bR^3}\left({\ts\prod\limits_x}\;\d^3\varphi_x\right)\sum_{m_x\in\bZ}\;\;\frac{2j+1}{4\pi}\int\limits_{S^2} \d n} \nonumber \\
&& \times\,
\exp\left[\sum_x\left(
-\frac{\beta}{16}\,\omega^2_{x\mu\nu} 
+ \frac{\irm}{2}\,\bb_{x\mu\nu}\cdot\omega_{x\mu\nu} 
- \frac{\irm}{2}\,\epsilon_{\rho\mu\nu}\nablab_\rho\varphi\cdot\omega_{x\mu\nu}  
+ 4\pi\irm\,|\varphi_x|\,m_x
\right)\right] \nonumber \\
&&
\eea
Here, $\bb_{x\mu\nu}$ stands for the $\bR^3$-valued 2-chain
\be
\label{definitionofbb}
\bb_{x\mu\nu} = (j+1/2)\,n\,S_{x\mu\nu}\,,
\ee
i.e.\ $\bb$ is proportional to the tensor product of the 2-chain $S_{x\mu\nu}$ (defining the surface $S$) and the unit vector $n$ in the Lie algebra vector space $\bR^3$.
We can express it also as
\be
\bb_{x\mu\nu} = -u_\mu\,(u\cdot\nabla)^{-1} J_{x\nu} + u_\nu\,(u\cdot\nabla)^{-1} J_{x\mu}
\ee 
where $u$ is the unit vector in the $x_1$-direction and the current is defined by 
\be
J_{x\mu} := (j+1/2)\,n\,C_{x\mu}\,.
\ee
Next we decompactify $\omega_{x\mu\nu}$ and integrate over it:
\bea
\lefteqn{\b \tr_j W_C\ket =
\frac{1}{Z}\,
\int\limits_{\bR^3}\left({\ts\prod\limits_x}\;\d^3\varphi_x\right)\sum_{m_x\in\bZ}\;\;\frac{2j+1}{4\pi}\int\limits_{S^2} \d n} \nonumber \\
&& \times\,
\exp\left[\sum_x\left(
-\frac{2}{\beta}\left(\nablab_\mu\varphi + \bb_{x\mu}\right)^2
+ 4\pi\irm\,|\varphi_x|\,m_x
\right)\right] 
\label{beforefactoringoff}
\eea
In this expression, we switched from the 2-chain $\bb_{x\mu\nu}$ to the 1-chain
\be
\bb_{x\rho} = \frac{1}{2}\,\epsilon_{\rho\mu\nu} \bb_{x\mu\nu}\,.
\ee
As in U(1) lattice gauge theory, we can factor off the Coulomb energy by making a change of variables and using the identity\footnote{Eq.\ \eq{identityforJ} can be proven by starting from the expression $\epsilon_{\rho\mu\nu}\nabla_\mu\bb^a_{x\mu}\Delta^{-1}\epsilon_{\rho\kappa\lambda}\nabla_\kappa\bb^a_{x\lambda}$ and observing that $J$ has no divergence.}
\be
\label{identityforJ}
\nabla_\mu\bb^a_{x\mu}\Delta^{-1}\nabla_\mu\bb^a_{x\mu} + \bb^2_{x\mu} = - J^a_{x\mu}\Delta^{-1}J^a_{x\mu}\,.
\ee
The final result reads
\bea
\lefteqn{\b \tr_j W_C\ket =
\frac{1}{Z}\,
\int\limits_{\bR^3}\left({\ts\prod\limits_x}\;\d^3\varphi_x\right)\sum_{m_x\in\bZ}\;\;\frac{2j+1}{4\pi}\int\limits_{S^2} \d n} \nonumber \\
&& \times\,
\exp\left[\sum_x\left(
\frac{2}{\beta}\,\varphi^a_x\Delta\varphi^a_x  
+ 4\pi\irm\left|\varphi_x + \Delta^{-1}\nabla_\mu\bb_{x\mu}\right|m_x
+ \frac{2}{\beta}\,J^a_{x\mu}\Delta^{-1}J^a_{x\mu}
\right)\right]
\label{gluonmonopolerepresentation}
\eea
It should be kept in mind that $\bb_{x\mu\nu}$ depends on the direction of the unit vector $n$ via eq.\ \eq{definitionofbb}.

We propose \eq{gluonmonopolerepresentation} as a non-abelian generalization of the photon-monopole representation.
The field $\varphi$ is interpreted as a dual gluon field: it has 3 degrees of freedom per point, which agrees with the fact that in 3 dimensions we have 1 physical degree of freedom per gluon and altogether 3 gluons for SU(2). It mediates the Coulomb interaction 
\be
V_{JJ} := -\frac{2}{\beta}\,\sum_x\;J^a_{x\mu}\Delta^{-1}J^a_{x\mu} = -\frac{1}{2}\,a g^2 (j+1/2)^2\sum_{xy}\;C_{x\mu}\Delta^{-1}_{xy} C_{y\mu}\,,\quad j\neq 0\,.
\ee
The latter agrees roughly with the tree-level result of standard perturbation theory \cite{Peterfulltwoloop,Petertothreeloop}: there one would have
\be
V^{\mathrm{tree}}_{JJ} = -\frac{1}{2}\,a g^2 j(j+1)\sum_{xy}\; C_{x\mu}\Delta^{-1}_{xy} C_{y\mu}\,.
\ee
The discrete variables $m$ arise from the compactness of SU(2) and can be seen as a generalization of monopoles to 3-dimensional SU(2) lattice gauge theory.

\section{Derivation of the area law}
\label{derivationofthearealaw}

The close similarity between the photon-monopole representation and \eq{gluonmonopolerepresentation} suggests that it could provide a way to generalize the derivation of confinement to 3d SU(2) lattice gauge theory. The main difference consists in the fact that the dual gluon field $\varphi$ is $\bR^3$- and not $\bR$-valued and that there appears a modulus in the second term. This renders the field theory nonlinear. As a result, we cannot perform a Gaussian integration to write down the analogue of a Coulomb gas representation.

In the abelian case, one can go to the Coulomb gas representation, extract a monopole self-energy and then transform back to an effective photon-monopole theory with a lower cutoff scale $M$.\footnote{see the appendix of ref.\ \cite{BanksMyersonKogut} and the clarifying remarks in 
\cite{DuncanMawhinney} and \cite{GopfertMack}.} The explicit appearance of the monopole self-interaction is important for deriving confinement, since it dampens monopole excitations.
 
In the present case, the nonlinearity forbids a simple transition to a Coulomb gas, and we proceed instead as follows: we will show that confinement can be derived \textit{if} the monopole self-energy behaves similarly as for U(1). We will assume that after renormalization down to a suitable cutoff scale $M$, 
\begin{itemize}
\item the field $\varphi$ is replaced by an effective field with a regulated Laplace operator, and 
\item a self-energy for the monopoles is generated.
\end{itemize}
If this is true, we will get
\bea
\lefteqn{\b \tr_j W_C\ket =
\frac{1}{Z}\,
\int\limits_{\bR^3}\left({\ts\prod\limits_x}\;\d^3\varphi_x\right)\sum_{m_x\in\bZ}\;\;\frac{2j+1}{4\pi}\int\limits_{S^2} \d n} \nonumber \\
&& \times\,
\exp\left[\sum_x\left(
\frac{2}{\beta}\,\varphi^a_x\tDel\varphi^a_x  
- 2\pi^2\beta v_0\,m^2_x 
+ 4\pi\irm\left|\varphi_x + \Delta^{-1}\nabla_\mu\bb_{x\mu}\right|m_x
+ \frac{2}{\beta}\,J^a_{x\mu}\Delta^{-1}J^a_{x\mu}
\right)\right]\,, \nonumber \\
&&
\eea
where the regulated Laplace operator is defined by
\be
-\tDel^{-1} := -\Delta^{-1} - \left(-\Delta + M^2\right)^{-1}\,,
\ee
and $v_0$ is some constant. From here on one can follow Polyakov's arguments to derive the area law \cite{PolyakovII}: let us abbreviate 
\be
\label{abbreviationetaSU2}
\eta_x = 4\pi\Delta^{-1}\nabla_\mu\bb_{x\mu}\,,
\ee
and apply a rescaling and shift on $\varphi$:
\bea
\lefteqn{\b \tr_j W_C\ket =
\frac{1}{Z}\,\int\limits_{\bR^3}\left({\ts\prod\limits_x}\;\d^3\varphi_x\right)\sum_{m_x}\;\;\frac{2j+1}{4\pi}\int\limits_{S^2} \d n\;\;\e^{-V_{JJ}}} \nonumber \\
&& \times\,
\exp\left[\sum_x\left(\frac{1}{8\pi^2\beta}\,(\varphi_x-\eta_x)\tDel(\varphi_x-\eta_x)
- 2\pi^2\beta v_0\,m^2_x 
+ \irm\,m_x|\varphi_x|
\right)\right]
\eea
Then, the dilute gas approximation yields
\bea
\label{SU2afterdilutegasapproximation}
\lefteqn{\b \tr_j W_C\ket =
\frac{1}{Z}\,
\int\limits_{\bR^3}\left({\ts\prod\limits_x}\;\d^3\varphi_x\right)\;\frac{2j+1}{4\pi}\int\limits_{S^2} \d n\;\;\e^{- V_{JJ}}} \nonumber \\ 
&& \times\,
\exp\left[\sum_x\left(
\frac{1}{8\pi^2\beta}\,(\varphi_x-\eta_x)\tDel(\varphi_x-\eta_x)
+ 2\,\e^{-2\pi^2\beta v_0}\cos|\varphi_x|
\right)\right]\,.
\eea
Recall that
\be
\label{recalldefinitionbb}
\bb_{x\mu\nu} = (j+1/2)\,n\,S_{x\mu\nu}\,,
\ee
and observe that the action is invariant under global rotations
\be
\varphi'{}^a = R^a{}_c\varphi^c\,,\qquad \bb'{}^a_{\!\!x\mu} = R^a{}_c\bb^c_{x\mu}\,.
\ee
The saddle points are determined by the equation
\be
-\tilde{\Delta} (\varphi-\eta)^a = M_D^2\frac{\varphi^a}{|\varphi|}\sin|\varphi|
\ee
where
\be
M_D^2 = 8\pi^2\beta\,\e^{-2\pi^2\beta v_0}\,.
\ee
We replace the regulated Laplacian by the full Laplacian, and plug in eq.\ \eq{abbreviationetaSU2}:
\be
\label{theequation}
\Delta\varphi^a = 4\pi\nabla_\mu\bb^a_{x\mu} - M_D^2\frac{\varphi^a}{|\varphi|}\sin|\varphi|
\ee
For simplicity, we assume now that $j=1/2$. Consider first the equation in the region above or below the surface $S$:
\be
\Delta\varphi^a = - M_D^2\frac{\varphi^a}{|\varphi|}\sin|\varphi|
\ee
We can find a simple solution for this if we assume that the direction of $\varphi$ is constant. Then the equation reduces to
the nonlinear Debye equation for $|\varphi|$.
\be
\label{equationwithoutsource}
\Delta |\varphi| = - M_D^2\sin|\varphi|\,,
\ee
We treat this as a quasi 1-dimensional problem and approximate it by the continuum equation
\be
\label{onedimensionalapproximation}
\pdiff{^2|\varphi|}{x^2_3} = -M_D^2\sin|\varphi|\,.
\ee
The term $4\pi\nabla_\mu\bb^a_{x\mu}$ in \eq{theequation} is only nonzero at the surface $S$ and gives
\be
-4\pi\nabla_\mu\bb^a_{x\mu} = 4\pi\,(\delta_{x_3,0} - \delta_{x_3,-a}) \,n^a\,.
\ee
This implies that at $S$ the field value $\varphi$ has to make a jump by $4\pi\,n$. 

If the jump was $2\pi\,n$, we could construct the solution as in the abelian case for charge $q = 1$.
The fact that it is $4\pi\,n$ creates a slight (but harmless) complication, and can be treated 
like the doubly charged loop of U(1) \cite{AmbjornGreensite}.

Recall that we have a certain freedom in choosing the particular solution $\bb_{x\mu\nu}$.
Instead of using the minimal surface, so that
\be
\bb = n\,S\,,
\ee
we could take two surfaces $S_+$ and $S_-$ s.t.\ $\pa S_+ = \pa S_- = C$, and set
\be
\bb_{x\mu\nu} = \frac{1}{2}\,n\,\left(S_+ + S_-\right)\,.
\ee
Now the $\varphi$-field has to jump two times by $2\pi\,n$: once along $S_-$, and a second time along $S_+$.
Thus, we can treat the situation near each surface similarly as for charge $q=1$.

Imagine that $S_+$ results from ``stretching'' the minimal surface $S$ to an $x_3$-value $\xb_3 > 0$,
and that $S_-$ is the mirror image of $S_+$ w.r.t.\ to the $x_1$-$x_2$-plane. Then, a solution
to \eq{theequation} is approximatively given by two domain walls: namely, 
\renewcommand{\arraystretch}{1.7}
\be
\varphi^{\mathrm{cl}}(x) \quad\approx\quad 
\left\{
\parbox{9.5cm}{
$\begin{array}{ll}
4\,n\,\arctan\left(\e^{-M_D (x_3+\xb_3)}\right) - 2\pi\,, & (x_1,x_2)\in S\,,\quad x_3 < 0\,,\\ 
4\,n\,\arctan\left(\e^{-M_D (x_3-\xb_3)}\right)\,, & (x_1,x_2)\in S\,,\quad x_3 > 0\,,\\ 
0\,, & \mbox{otherwise}\,.
\end{array}$
}\right.
\ee
We plug this solution back into \eq{SU2afterdilutegasapproximation} and use the trivial saddle point for $Z$:
\be
\b \tr_j W_C\ket =
\frac{1}{4\pi}\int\limits_{S^2} \d n\;\;\e^{- V_{JJ}}
\exp\left[\sum_x\left(
\frac{1}{8\pi^2\beta}\,(\varphi^{\mathrm{cl}}_x-\eta_x)\tDel(\varphi^{\mathrm{cl}}_x-\eta_x)
+ 2\,\e^{-2\pi^2\beta v_0}\cos|\varphi^{\mathrm{cl}}_x|
\right)\right] 
\ee
The associated action equals $A \clS_{\mathrm{2dw}}$, where $A$ is the area of the surface $S$ and $\clS_{\mathrm{2dw}}$ is the action (or minus the energy) of
the two domain walls, i.e.\
\be
\clS_{\mathrm{2dw}} \approx \frac{4 M_D}{\pi^2\beta}\,.
\ee
The action does not depend on $n$, so the integral over $n$ drops out. The result is an area law
\be
\b \tr_j W_C\ket\quad\approx\quad 2\,\exp\left(- V_{JJ} - \sigma A\right)
\ee
with a string tension
\be
\sigma = \frac{4 M_D}{\pi^2\beta}\,. 
\ee
As in the case of the Coulomb potential, it would be wrong to set $j$ to zero in formula \eq{recalldefinitionbb}.
This would yield a confining potential when there is actually no Wilson loop.
The reason is that we use the Kirillov trace formula in conjunction with a stationary phase approximation,  
and that creates a wrong offset for the $j$-dependence.
The correct procedure is to set $j=0$ at the beginning of the derivation. Then, the resulting potential is zero, as it should.
\renewcommand{\arraystretch}{1}

\section{Summary and discussion}
\label{summaryanddiscussion}

In this paper, we have derived an approximative model for 3-dimensional SU(2) lattice gauge theory. Its degrees of freedom can be viewed as 
a dual gluon and monopole field. We propose it as a generalization of the photon-monopole representation of Polyakov \cite{PolyakovI,PolyakovII} and Banks, Myerson and Kogut \cite{BanksMyersonKogut}.

Our derivation extends an earlier work by Borisenko, Voloshin and Faber \cite{BorisenkoVoloshinFaber}, where a weak-coupling perturbation theory of the plaquette representation was developed. We only retain the zeroth order of this expansion, but include a non-perturbative contribution from a Wilson loop. 
Thus, our method can be seen as a stationary phase approximation around non-trivial field configurations: these are given by monopoles that arise from the compactness of the gauge group, and by a worldsheet that is bounded by the Wilson loop.

The resulting model contains two interaction terms: firstly, a current-current potential that is essentially the tree-level Coulomb interaction one would get from a purely perturbative treatment. Secondly, a coupling between monopoles, dual gluons and current. This coupling is similar to the photon-monopole coupling of U(1), but nonlinear.  

The analogy with U(1) suggests a possible derivation of confinement. It requires us, however, to make an additional assumption on the monopole self-energy.
In the abelian case, one can go to the Coulomb gas representation, extract a monopole self-energy and transform back to an effective photon-monopole representation. Here, we cannot compute the analogue of a Coulomb gas, since the integral over the dual gluon field is non-Gaussian.
Instead we \textit{assumed} that a renormalization generates a monopole self-energy at a lower energy scale $M$.
From there on, we can proceed as in Polykov's derivation and arrive at a non-vanishing string tension for the Wilson loop.
Further investigation has to show if our assumption is justified.

In the two companion papers, we start from two other representations of 3d SU(2) lattice gauge theory and arrive at models that are quite similar to the present one. 

Whether the model of this paper is a good or a bad approximation can be tested: it is simpler than the full lattice gauge theory and may be easily implemented on a computer. By summing over the monopole variables in expression \eq{beforefactoringoff} one can remove the phase factors and translate them into a constraint. The latter can be enforced by a Gaussian damping factor. 

\section*{Acknowledgements}

I thank Abhay Ashtekar, Gerhard Mack, Alejandro Perez and Hendryk Pfeiffer for discussions. 
This work was supported in part by the NSF grant PHY-0456913 and the Eberly research funds. 

\bibliography{bibliography}
\bibliographystyle{hunsrt}  

\end{document}